# ROOT Status and Future Developments

R. Brun, F. Rademakers
*CERN, CH 1211, Geneva 23, Switzerland*

P. Canal
*FNAL, Batavia, IL 60510, USA*

M. Goto
*Agilent Technologies, Suginami-ku, Tokyo 168,  JAPAN*

In this talk we will review the major additions and improvements made to the ROOT system in the last 18 months and present our plans for future developments. The additions and improvements range from modifications to the I/O sub-system to allow users to save and restore objects of classes that have not been instrumented by special ROOT macros, to the addition of a geometry package designed for building, browsing, tracking and visualizing detector geometries. Other improvements include enhancements to the quick analysis sub-system (TTree::Draw()), the addition of classes that allow inter-file object references (TRef, TRefArray), better support for templates and STL classes, amelioration of the Automatic Script Compiler and the incorporation of new fitting and mathematical tools. Efforts have also been made to increase the modularity of the ROOT system with the introduction of more abstract interfaces and the development of a plug-in manager. In the near future, we intend to continue the development of PROOF and its interfacing with GRID environments. We plan on providing an interface between Geant3, Geant4 and Fluka and the new geometry package. The ROOT GUI classes will finally be available on Windows and we plan to release a GUI inspector and builder. In the last year, ROOT has drawn the endorsement of additional experiments and institutions. It is now officially supported by CERN and used as key I/O component by the LCG project.

## 1. INTRODUCTION

Since its inception in 1995 by René Brun and Fons Rademakers, ROOT has gone through several releases and has gained wide acceptance in the HEP community, in the research communities in general and even in several for profit organizations.  ROOT has been ported to more that 40 platform, OS and compiler combinations and is currently released in binary format for more than 30 Platforms.

## 2. OFFICIAL SUPPORT

In 2003, CERN has decided to officially support the ROOT project.  The support will be provided by the SFT group in the EP division.  In concrete term, this results in additional manpower for the project.  This manpower will in particular benefit the geometry, graphics and documentation part of the project.

### 2.1. Current Team Members and Associates.

ROOT Team:
- Ilka Antcheva, CERN
- Rene Brun, CERN
- Philippe Canal, FNAL
- Olivier Couet, CERN
- Gerardo Ganis, CERN
- Masa Goto, Agilent technologies
- Valeriy Onuchin, CERN
- Fons Rademakers, CERN

Associates:
- Bertrand Bellenot (WinGdk), private
- Maarten Ballintijn (PROOF), MIT/Phobos
- Andrei Gheata: (Geometry package), CERN/Alice
- Valery Fine (TVirtualX/Qt, I/O), BNL/STAR/Atlas
- Victor Perevoztchikov (STL, foreign classes), BNL/STAR /Atlas

And more than 50 other important contributions have been made by people spending a substantial fraction of their time on the project.
See $ROOTSYS/README/CREDITS

Special thanks to Suzanne Panacek who did a great job with the ROOT Users Guide, tutorials, lectures.

Many thanks to FNAL computing Division for the continuous support of the project since 1998

## 3. DEVELOPMENTS SINCE CHEP 2001

### 3.1. ROOT I/O developments

#### 3.1.1. Foreign Classes

We added the ability to store in a ROOT file, objects of classes that have not been instrumented for ROOT in any way.  To enable this feature, the user just need to generate, compile and link a ROOT dictionary file for the classes she wants to store in a file.  This can be particularly useful for 3$^{rd}$ party libraries where you are not able or not allowed to modify the header files or source code.

#### 3.1.2. Emulated Classes

The ROOT file format has been upgraded to be self-describing in the default cases.  When the objects of a class are saved in a ROOT file using the "new I/O style", a description of how these objects were saved is also saved in the file. ROOT can re-use this information to read the file even if the original library or code is not available.

We use the term "Emulated Classes" to describe classes that are read using only the self-describing information stored with the file.

### 3.1.3. TRef and TRefArray

The TRef and TRefArray classes are designed to provide a lightweight implementation of a persistent link with very fast dereferencing.

Given an object A and an object B who both points to an object C and given that this link is implemented using a bare C++ pointer (or equivalent), if A and B are saved in 2 different buffers (i.e. 2 different I/O operations), the object C will be duplicated on file. Worse, when reading back the object A and B, 2 distinct objects C will be created.

To avoid this duplication, one of the bare C++ pointers can be replaced by a TRef (or part of a TRefArray) object. In this case, instead of saving the object C a $2^{nd}$ time, a TRef object is saved with enough information to be able to reconnect to the object C when reading back. The TRef object and the object C can be saved in 2 different files; it will be the user responsibility to make sure that the object C has been read when it is needed. If the C has not yet been read and the TRef is access, the TRef will return a null pointer. Once the object C is read, any further access to the TRef object will return the actual address of C.

### 3.1.4. TTree

Several improvements were made to the TTree class.

We added the ability to create branches directly from a collection of objects. In particular, this allows the users to avoid hard-coding the list of objects placed in the file.

We improved the split algorithm to support many additional complex cases of inheritance and composition.

We added an automatic file overflow. When the file in which the TTree object is being stored reaches a user specified maximum size, the file is automatic switch over. This means that a new file is created and the histograms and trees that were being saved in the original file are now saved in the newly opened file.

If the original file name was "myfile.root", the new file will be named "myfile_1.root" and then "myfile_2.root"

### 3.1.5. Histograms

We added a new class called THStack. A THStack is a collection of TH1 (or derived) objects. By default (if the option "nostack" is not specified), the histograms will be painted stacked on top of each other.

Many other functions and drawing options were added to TH1 (see the documentation on the web [6]).

We added the ability to histogram 'strings'.

We added the ability to merge a collection of histograms.

## 3.2. Geometrical modeler

Geometrical modeling generally provides the geometrical description of a device and a set of services to "navigate" through its structure. HEP geometrical modelers are in particular designed to handle high complexity detector geometries and they are usually embedded within Monte Carlo (MC) simulation frameworks. The fact that these frameworks greatly depend on their specific geometrical tools makes simulation applications hardly portable to MC's other than the one they were designed for.

The ALICE off-line project in collaboration with the ROOT team is developing a multi-purpose geometrical modeler for HEP that is integrated within a virtual MC scheme. This tool has been optimized for performance with all the geometry setups of LHC experiments and provides a unique representation for the geometry used by applications such as simulation, reconstruction or event display [1].

Performance has been the highest priority during the development and this is reflected by the benchmarks. The code is now available in the ROOT distribution.

## 3.3. PROOF

The Parallel ROOT Facility, PROOF, is designed for the interactive analysis of very large sets of ROOT data files on a cluster of computers. PROOF enables a physicist to analyze and understand much larger data sets on a shorter time scale. It makes use of the inherent parallelism in event data and implements an architecture that optimizes I/O and CPU utilization in heterogeneous clusters with distributed storage. The system provides transparent and interactive access to gigabytes today. Being part of the ROOT framework PROOF inherits the benefits of an efficient Object storage system and a wealth of statistical and visualization tools.

The main idea is to speed up the query processing by employing parallelism. This is achieved by using a three-tier architecture to distribute the user analysis code to a set of worker nodes that then ask a master node for data to be processed (pull architecture). The master optimizes the work distribution to the worker nodes in function of data locality and worker response. In the GRID context, this model will be extended from local cluster to a wide area "virtual cluster", where the worker nodes are started at different sites under the control of a GRID resource broker. The emphasis in that case is not so much on interactive response as on transparency. With a single query, a user can analyze a globally distributed data set and get back a single result [2], [3].

### 3.4. GRID And ROOT

PROOF has been designed so that it would be able to make use of the GRID tools. It will be able to use Grid Resource Brokers, Grid File Catalogs and Grid Monitoring Services in order to discover the best resources to use when parallelizing a job. In particular, PROOF will be able to connect with the CONDOR tools.

To enable this link to the GRID services in as transparent a way as possible, we added the class TGrid, an abstract interface to the GRID services. A first concrete implementation (TAlien) was developed by P. Buncic, A. Peters, P. Saiz for the ALICE collaboration [4].

The ROOT file access framework (TFile, TDirectory, etc.) was designed to allow the easy addition of support for new I/O protocols. This features was used to add support for the following protocols and servers:
- CHIRP, the remote I/O protocol from CONDOR
- DCache a distributed random access mass storage cache developed by DESY and Fermilab
- RFIO, the remote I/O protocol used at CERN for Castor

### 3.5. Graphics Improvements

Several additions were made to the Graphics capabilities of ROOT.

You can now draw TF3 objects:

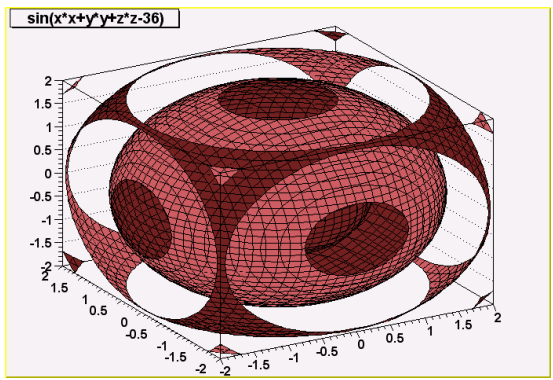

ROOT is now able to produce SVG files using the TSVG class. TSVG may be used like TPostScript to produce a Scalable Vector Graphics file instead of a postscript file. Viewers like Internet Explorer can view directly the SVG files.

Thanks to Christian Stratowa, ROOT now has a class TGraphSmooth that implement the smoothing or TGraph, TGraphErrors and the interpolation a graph at a set of given points. See the new tutorial motorcycle.C.

Using an external library called AfterImage, ROOT now offers many new 2D graphics capabilities. In particular those new capabilities are used to render astronomical images:

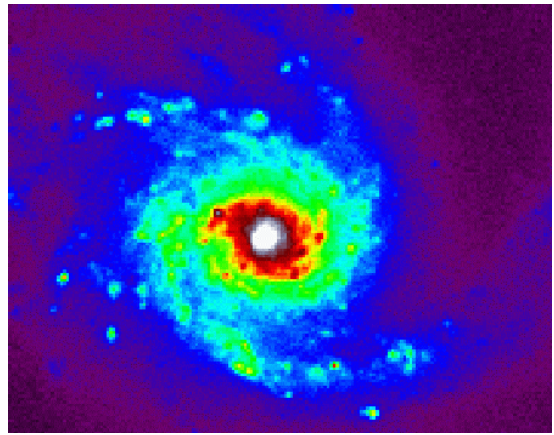

### 3.6. MS. Windows Support

Thanks to Bertrand Bellenot, a full port of the new ROOT graphical widgets to the Windows operating system is now available (wingdk) and will soon be made the default [5].

All the widgets are now fully functioning. They offer the same look and feel on Windows and on Unix and are fully compatible between the 2 platforms. OpenGL is also supported on both platforms.

However, the port on windows is still slower than the previous interface. Sockets and memory-mapped files still need to be ported.

The next priority will be to implement the sockets and to improve the speed.

Once this is accomplished memory mapped files and X3D will also be ported.

In the longer run, we would like to implement rootd as a Windows service.

### 3.7. Other New Features And Classes

TMatrix is now actively supported and developed by Eddy Offermann from Renaissance Technologies. He has already made many enhancements.

Several new functions were added to TMath, including the Bessel functions, Voigt function, BreitWigner function, Struve functions, IsInside and a couple of new sort algorithms (Adrian Bevan). Tony Colley added many fundamental and physical constants.

Frank Filthaut contributed the TFractionFitter class, which can be used to fit Monte-Carlo fractions to the histogram data, taking in consideration both the data and the Monte Carlo statistical uncertainties.

Christophe Delaere contributed a set of classes to handle Confidence Levels (TLimit, TLimitDataSource and TConfidenceLevel).

Adrian Bevan also contributed a class to calculate the Confidence Level upper limit using the Feldman-Cousins method.

As we can see ROOT's growth benefits more and more from the contributions from its users, including those from for-profit organizations.

### 3.8. System Enhancements

To enhance the modularity and flexibility of the ROOT framework, a Plug-in Manager (TPluginManager) was introduced to remove hard dependences on plug-ins.

This allows easy extension of abstract interface. In the configuration file, you can list
- base class name
- regular expression to recognize this entry
- plug-in class name
- plug-in library name
- constructor prototype

For example:

```
# base class    regexp   plugin class    plugin lib
  ctor or factory
Plugin.TFile:  ^rfio:    TRFIOFile       RFIO
   "TRFIOFile(const char*,Option_t*,const char*,Int_t)"
+Plugin.TFile: ^dcache:TDCacheFile  DCache
   "TDCacheFile(const char*,Option_t*,const char*,Int_t)"
```

ACLiC now checks all the files that are included in a script before deciding whether or not it should be recompiled. In particular this means that we recommend using the 'refresh as necessary option ('+') rather than the 'always recompile option' ('++').

ACLiC now allows the user to select whether the library should be compiled in debug or optimized mode.

It also supports the compilation of scripts stored in a read-only directory and the ability to store the libraries in a user defined location.

Rootcint (and thus ACLiC) is now properly handling CINT's pragma statements anywhere in the header files. Previously these statements had to be stored in a linkdef file.

Cint and Rootcint has also been improved significantly, in particular for
- Class templates
- STL containers
- I/O for 'foreign classes'
- Classes with multiple-inheritances.

### 3.9. Port To New Platforms

ROOT has been ported to following new platforms:
- MacOS
  thanks to Ben Cowan, Keisuke Fujii, George Irwin, and al.
- Windows with cygwin and gcc 3.2
  thanks to Axel Naumann
- Intel Compiler v7
- Itanium 64

## 4. CURRENT DEVELOPMENTS

In the following sections we describe the features and enhancements that are likely to be introduced in ROOT in the next few months.

### 4.1. ROOT I/O

We plan to lift the remaining limitations that still exist for the I/O of the foreign classes. Those restrictions are mainly on the Windows platform in case of complex nesting of private classes.

We will introduce a new implementation of the STL container I/O mechanisms which will allows:
- Splitting of std::vector and std::list
- Emulation of STL container classes
- TTree::Draw'ing of STL containers. And their content.

We plan to introduce full support for very large file (more that 2 Gb) on platform that support such large files.

We intent to introduce an XML interchange format, where instead of saving an object to a ROOT file, the user will be able to save it to an ASCII, xml formatted file. This should be done in such a way that switch from one representation to the other will represent minimal effort on the user's part.

### 4.2. TTree

As mentioned in the previous paragraph, we plan to introduce support for splitting STL lists and vectors in a fashion similar to the splitting of TClonesArray objects.

A new feature in TTree::Draw will allow the user to provide a function that will be executed for each entry of the TTree in a context such that the branch name can be used to read the values stored in the branch:
- Branches will be read only if accessed
- The function can use arbitrary C++
- The return value of the function will be histogrammed

This will work in both interpreted and compiled mode. Additionally there will be a new generation of the MakeClass utility that will generate stubs using the same technology.

TTree::Draw will also be updated so that it can call any numerical global function or static member functions.

### 4.3. New Graphical Interface

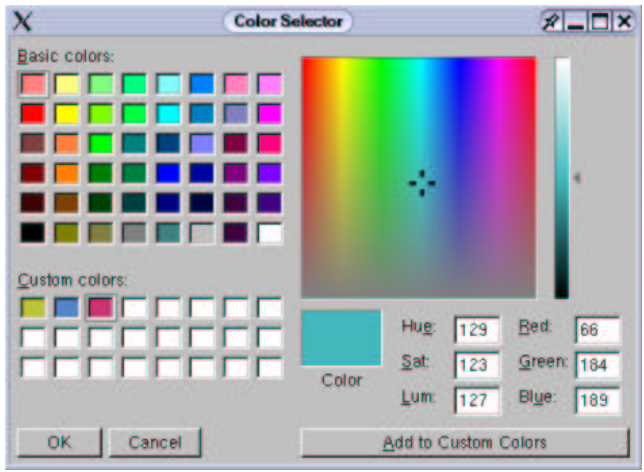

One of the features of the default graphical interface of ROOT is to be totally cross-platform. Due to this feature and the fact that the newest widgets were not ported to Windows, the graphical interface has not been updated in a long time. Now that the port of the widget to Windows is nearly completed, we are now able to redesign the existing Graphical User Interface. The intent is to bring the core panels more in line with the modern interfaces.

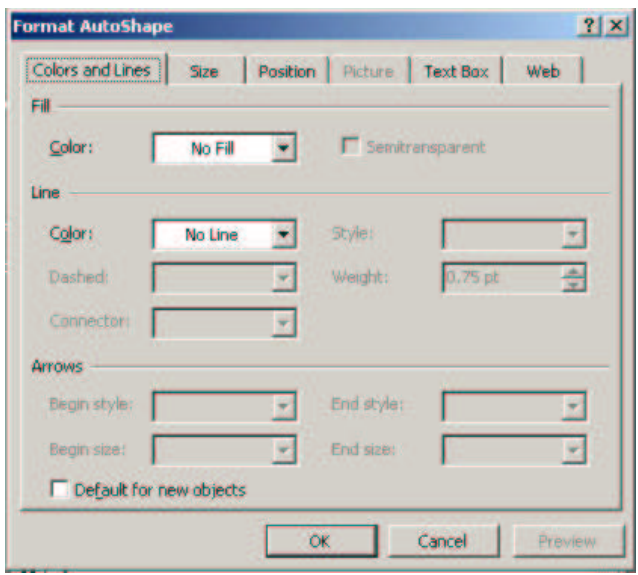

In addition we plan on develop a ROOT graphical interface builder as well as releasing an addition to the ROOT User's Guide documentation the ROOT GUI Classes.

We are also intending on releasing, with the help of Valery Fine, a QT implementation of the ROOT graphical interface classes.

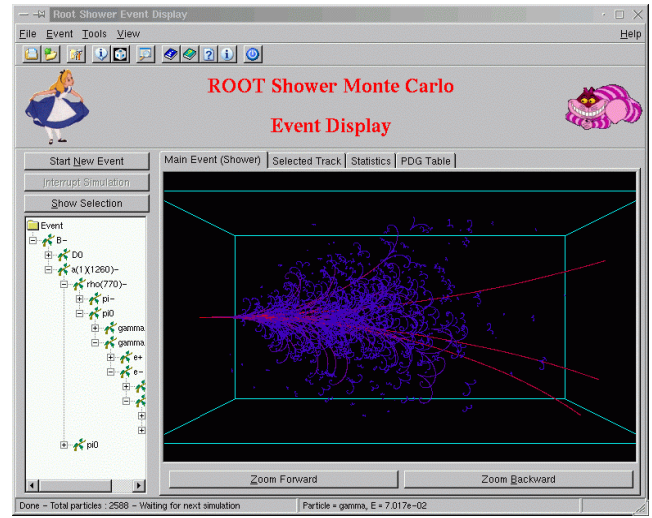

### 4.4. New online help system

ROOT already has the ability, via the THtml class, to produce a set of web pages documenting any C++ classes. We would like to extend this model and make the same information available directly in a ROOT process.

To do, we would produce a new file containing the documentation for a specific library in a very compact form. We will also add the interface, both programmatic and graphical to access quickly this information.

### 5. CONCLUSIONS

Even-though ROOT has now reached a level of stability and functionality that makes it the ideal platform for large data set storage and analysis; it is still under very active development, not only by the original developers but also thanks to the many contributions of its users.

Additional up to date information can be found on the ROOT web pages [6].

### References

[1] R. Brun, A. Gheata, M. Gheata, A Geometrical Modeler, Presented at CHEP'03, La Jolla, PSN THMT001

[2] M.Ballintijn, R.Brun, F.Rademakers and G.Roland, Distributed Parallel Analysis Framework with PROOF, Presented at CHEP'03, La Jolla, PSN TUCT004

[3] M.Ballintijn, R.Brun, F.Rademakers and G.Roland, Analyse your Data in Parallel using PROOF, Presented at CHEP'03, La Jolla, PSN TULT003

[4] P.Buncic, A.Peters, P.Saiz, The AliEn system, status and perspectives, Presented at CHEP'03, La Jolla, PSN THAT005.

[5] Status of Native ROOT GUI Port to Win32: http://root.cern.ch/root/win32progress/Win32GUI.html

[6] http://root.cern.ch